\documentclass{svjour3}                     
\smartqed  
\usepackage{graphicx}
\usepackage{mathptmx}      
\usepackage{natbib}
\usepackage{multicol}
\newcommand{\hone}{H~{\footnotesize{I}}}  	
\newcommand{\carthree}{C~{\footnotesize{III}}}  
\newcommand{\oxysix}{O~{\footnotesize{VI}}}  	
\newcommand{\oxyseven}{O~{\footnotesize{VII}}}  
\newcommand{\oxyeight}{O~{\footnotesize{VIII}}} 
\newcommand{\ironeight}{Fe~{\footnotesize{VIII}}} 
\newcommand{\ironeleven}{Fe~{\footnotesize{XI}}} 
\newcommand{\chandra}{{\it{Chandra}}}		
\newcommand{\chips}{{\it{CHIPS}}}		
\newcommand{\copernicus}{{\it{Copernicus}}}	
\newcommand{\euve}{{\it{EUVE}}}			
\newcommand{\fuse}{{\it{FUSE}}}			
\newcommand{\rosat}{{\it{ROSAT}}}		
\newcommand{\xmm}{{\it{XMM}}}                   
\newcommand{\suzaku}{{\it{Suzaku}}}             

\begin{document}

\title{Report from Sessions 1 and 3, including the Local Bubble Debate}

\titlerunning{The Local Bubble Debate}        

\author{Robin L. Shelton
\and
Members of Sessions 1 and 3}


\institute{Department of Physics and Astronomy, the University of Georgia,
        Athens, GA 30602}



\date{Received: date / Accepted: date}

\maketitle

\begin{abstract}

This report summarizes the discussions in the Session 1 and Session 3 groups  
which met to discuss the questions:
``What Physical Processes Drive the Multiphase Interstellar Medium 
in the Local Bubble?'', and 
``What are the Energy and Pressure Balances in the Local Bubble?'' 
Most of our understanding of the Local Bubble has
come from soft X-ray observations, but recent appreciation of the importance 
of solar wind charge exchange (SWCX) reactions has shown that the heliosphere
produces some fraction of the soft X-rays that were previously ascribed 
to the Local Bubble.
Some astronomers suggest that the SWCX X-rays rather than Local Bubble
emission could explain most of 
the locally produced X-rays.   Our discussions, therefore,
also included a debate concerning the Local Bubble's existence.

\end{abstract}


\section{Introduction}
\label{intro}

The workshop conveners created small groups which were asked to discuss 
issues related to the heliosphere and the Local Bubble (LB).   The members of
Sessions 1 and 3 combined to discuss the physical processes within and
pressure and energy balances of the LB.
While these topics assume that the LB actually exists,
this assumption is debatable because
the traditional evidence for a hot LB (soft X-rays that originate within
a few hundred parsecs of the Earth and are seen in all
directions) could possibly be due to another, recently discovered 
source.
That source, solar wind charge exchange (SWCX) is outlined in 
Section ~\ref{swcx}.   
The debate over the Local Bubble's existence is presented in
Section~\ref{debate}.
In Section~\ref{pressure_energy}, we summarize the discussion of the pressure
and energy balances within the Local Bubble.  The discussion 
of the pressure balance allows for the possibility that the LB gas pressure
is smaller than previously assumed due to the SWCX component,
but the later discussions assume the traditional conception of the LB 
as containing hot gas.
In Section~\ref{smallstructure}, we summarize the discussion of 
clouds within the
Local Bubble, and in Section~\ref{otherprocesses}, we discuss
other important physical processes.   This report follows the arguments
presented at the session and later presented to the workshop participants 
for their comments.

\section{ Solar Wind Charge Exchange (SWCX)} 
\label{swcx}

\begin{enumerate}
\item
The solar wind is highly charged.   When solar wind ions
encounter neutral gas (in the coma of comets,
in planetary atmospheres,
and dispersed within the heliosphere),
they have a high probability
of capturing an electron from the neutral atom.

\item
An example using solar wind element ``X'': \hspace*{0.2cm}
X$^{(n+1)+}$ + H or He $\Rightarrow$ X$^{n+*}$ + H$^{+}$ or He$^{+}$.

\item
Since the electron is transferred into a highly excited orbital (denoted 
by $*$), the 
charge transfer is soon followed by de-excitation and radiation 
of at least one X-ray photon.   
The resulting X-rays are colloquially referred to as SWCX X-rays.

\item
Cravens (2000) and Robertson \& Cravens (2003) proposed that
$\sim50\%$ of the observed 1/4 keV flux 
previously attributed to the LB is due to SWCX in the heliosphere.

\item
In the 3/4 keV band, Koutroumpa et al. (2007) proposed that SWCX can
explain $\sim100\%$ of the observed 
\oxyseven\ X-ray flux previously attributed to the LB.

\item
Charge exchange and the subsequent release of X-ray photons
are being studied currently with more refined reaction rates 
expected in the near future.
\item
The SWCX intensity depends on the Sun's activity level, which varies with time.
\end{enumerate}

\section{ After We Consider the Effects of SWCX, Is There Reason
to Believe that the LB Actually Exists?   
\hspace{0.3cm}  
---
\hspace{0.3cm}  
the Local Bubble vs Solar Wind Charge Exchange Debate}
\label{debate} 

Below, we present the arguments for (left column) and against (right column)
the existence of the Local Bubble:   

 \begin{multicols}{2}    

\noindent
{\large\bf{Pro Local Bubble}} \\

\noindent
1.) {\large\it{X-Ray Argument:\ \  
SWCX explains only a fraction of the observed ``locally produced''
X-rays, the remainder must come from the Local Bubble:}} 
\begin{enumerate}
\renewcommand{\labelenumi}{\Alph{enumi}}
\item
Cravens (2000) and Robertson \& Cravens (2003) 
predicted that $\sim1/2$ of the diffuse 1/4 keV X-rays
observed in the Galactic plane by \rosat\ are from SWCX.
The remainder must be from the LB.
\begin{itemize}
\item[$\bullet$]
The Robertson \& Cravens predictions 
were made for the same
lines of sight through the heliosphere
as were used in the \rosat\ survey
\end{itemize}
\item
The distribution of SWCX 1/4 keV photons
(see map in Robertson \& Cravens (2003))
differs significantly from the observed
distribution (see map in Snowden et al. 
(1998)).
The difference must be due to the hot LB
contribution. \\
\end{enumerate}

\noindent
{\large\bf{Anti Local Bubble}} \\

\noindent
1.) {\large\it{X-Ray Argument: 
SWCX generates all of the ``locally produced''
3/4 keV X-rays, so there is no need to assume the existence
of a Local Bubble:}} 

\begin{enumerate}
\renewcommand{\labelenumi}{\Alph{enumi}}
\item
SWCX predictions for 2 sample LB observations explain 
all of the observed local \oxyseven\ and \oxyeight\ emission 
at 0.57 and 0.65 keV
(Koutroumpa et al. 2007).
\begin{itemize}
\item[$\bullet$]
These calculations were done for the same lines
of sight through the heliosphere as the
\chandra, \xmm, and \suzaku\ observations.
\item[$\bullet$]
These calculations included the SWCX enhancements associated
with coronal mass ejections.
\end{itemize}
\end{enumerate}
. \\
. \\
. \\
. \\
. \\
. \\
  \end{multicols}

\begin{multicols}{2}    

\noindent
{\large\bf{Pro Local Bubble, continued}} \\

\begin{enumerate}
\renewcommand{\labelenumi}{\Alph{enumi}}
\addtocounter{enumi}{2}
\item
SWCX intensities vary with time and direction, but
several soft X-ray surveys made during different years and using
different angles between the 
target and Sun found similar soft X-ray background 
fluxes.
\begin{itemize}
\item[$\bullet$]
The fluxes observed by the \rosat, SAS 3, and Wisconsin survey 
are shown to be correlated in 
Figure 7 of Snowden et al. (1995).
\end{itemize}
\item
The observed 1/4 keV X-ray flux is bright where the \hone\ column 
density is dim and {\em vice verse}.   This anticorrelation
implies that the X-ray emitting gas occupies a cavity in the
\hone\ distribution. In general, the effect is not due to absorption, which
would simultaneously harden the observed spectrum
(see McCammon \& Sanders, 1990).\\
\end{enumerate}

\noindent
{\large\bf{Anti Local Bubble, continued}} \\

\noindent
. \\
. \\
. \\
. \\
. \\
. \\
. \\
. \\
. \\
. \\
. \\
. \\
. \\
. \\
. \\
. \\
. \\

\end{multicols}
\begin{multicols}{2}

\noindent
2.) {\large\it{\oxysix\ Argument:
\oxysix, which traces $T = 3 x 10^5$~K gas at interfaces between hotter and 
cooler gas, has been observed, so there must be hot gas in the LB:}}

\begin{enumerate}
\renewcommand{\labelenumi}{\Alph{enumi}}
\item  Absorption by \oxysix\ ions has been seen on 
short sight lines that are within or extend somewhat
beyond the radius of the LB (Jenkins 1978; 
Oegerle et al. 2005; Savage \& Lehner 2006.) 
The existence of \oxysix\ implies the existence
of a hot LB.\\
\begin{itemize}
\item[$\bullet$]
Response:  Surely, not all of the \oxysix\ observations could be wrong: \\
Some of the \fuse\ observations used cool ($T < 40,000$~K) white
dwarf stars as background sources and the \copernicus\ observations 
used B stars. These stars should not have photospheric \oxysix.
\end{itemize}
\end{enumerate}

\noindent
2.) {\large\it{\oxysix\ Counter-Argument: 
Most of the observed \oxysix\ absorption features 
can be explained without a LB:}} 

\begin{enumerate}
\renewcommand{\labelenumi}{\Alph{enumi}}
\item
Many of the observed \oxysix\ absorption features can be
explained in other ways 
\begin{itemize}
\item[$\bullet$]
Savage \& Lehner used nearby white dwarfs.   Some of
their white dwarf spectra
were contaminated by stellar or circumstellar \oxysix\ absorption,
which Savage \& Lehner incorrectly attributed to the local ISM
(Barstow 2008; Welsh 2008).
\item[$\bullet$]
Savage \& Lehner processed their \fuse\ data using an old 
version (v2.5) of the FUSE pipeline;  
when a newer version was used, 
some of the \oxysix\ features disappeared (Barstow, 2008).
\end{itemize}
\end{enumerate}

\end{multicols}

\pagebreak

\begin{multicols}{2}

\noindent
3.) {\large\it{EUV Counter-Argument:  
The \chips\ and \euve\ measurements 
are ambiguous}} 
\begin{enumerate}
\renewcommand{\labelenumi}{\Alph{enumi}}
\item
Responses: 
\begin{itemize}
\item[$\bullet$]
The models assumed CIE
and solar abundances of gas-phase
iron.  Either assumption could be incorrect
\item[$\bullet$]
The \euve\ observations are thought to have been contaminated 
by a very soft X-ray leak. 
\item[$\bullet$]
The calorimeter (McCammon et al. 2002) saw a much larger
intensity in the $170$ \AA\ region than did \chips\ or
\euve.
\end{itemize}
\end{enumerate}

\noindent
3.) {\large\it{EUV Argument: 
LB models predict Extreme UV emission that 
was not seen by \euve\ or \chips\ }}
\begin{enumerate}
\renewcommand{\labelenumi}{\Alph{enumi}}
\item
\euve\ and \chips\ did not find as much
\ironeight\ to \ironeleven\ emission
around 170~\AA\
as expected from models of the hot LB
(Jelinsky et al. 1995; Vallerga \& Slavin, 1998;
Hurwitz et al. 2005).
\end{enumerate}
. \\
. \\

\end{multicols}

\begin{multicols}{2}

\noindent
4.) {\large\it{
Local Cavity and Local Cloud Argument
(also see Section~\ref{pressure_energy}):
}}

\begin{enumerate}
\renewcommand{\labelenumi}{\Alph{enumi}}
\item
The Local Cavity with little \hone\
(Lallement et al. 2003) 
must be filled with ionized gas (such as the LB), otherwise
it would collapse.   Similarly, the Local Interstellar Cloud
(see other articles in this volume) would
expand if it were not compressed by the
Local Bubble's pressure.
\begin{itemize}
\item[$\bullet$]
Response: Invoking a large magnetic pressure in the Local Cavity 
requires an unreasonably strong magnetic field.
\item[$\bullet$]
Response:  If the gas is not in pressure balance, then the Local Cloud 
would be expanding at ~10 km s$^{-1}$ which is inconsistent 
with observations.
\end{itemize}
\item Stars within the Local Bubble region have outflows that
terminate where they interact with the ISM.   Therefore, for
termination shocks to exist here, there must be material in the LB
(Wood 2007).
\end{enumerate}

\noindent
4.) {\large\it{Local Cavity and Local Cloud Counter-Argument: 
}}

\begin{enumerate}
\renewcommand{\labelenumi}{\Alph{enumi}}
\item
There is an unexplained gap between the LB and the Local Cavity wall, 
so the Milky Way does contain regions with apparent imbalances in 
the thermal pressure.
\begin{itemize}
\item[$\bullet$]
Magnetic pressure may compensate for the imbalance in thermal pressure.
\item[$\bullet$]
Total pressures need not be balanced if the system is dynamic.
\end{itemize}
. \\
. \\
. \\
. \\
. \\
. \\
. \\
. \\
. \\
. \\
. \\
\end{enumerate}

 \end{multicols}

\pagebreak

\subsection{ What measurements are needed to 
determine how much hot gas may exist in the Local Bubble? } 

\begin{enumerate}
\item
More X-ray spectroscopic observations are needed to test the
SWCX predictions. 
\item
X-ray observations with much higher spectral resolution 
are needed to compare the spectral
details of the SWCX models with observed spectra.
\item
The high spectral resolution observations must have sufficient spatial
resolution for shadowing experiments, in which opaque clouds are
used to block the X-rays from more distant diffuse sources.
\item
Such high spectral and spatial resolution observations 
are needed in the 1/4 keV band, which carries much of the flux
of the presumably hot ($T \sim 10^6$~K) Local Bubble.
\end{enumerate}

\section{Pressure and Energy Balances in the Local Bubble} 
\label{pressure_energy} 

\subsection{ Pressure Balance} 

\begin{enumerate}

\item
Previously, the
LB's thermal pressure had been calculated from the
flux of 1/4~keV photons produced between the Sun and a shadowing
cloud located $\sim 90$~pc from the Sun.
The resulting thermal pressure was
$P_{th}/k \sim 15,000$~K cm$^{-3}$ (Snowden et al. 1998).

\begin{enumerate}
  \item
The LB pressure quoted above is 
far larger than the thermal pressure of the local warm clouds and
other cool gas in the Galactic disk,
$P_{th}/k \sim 2200$~K cm$^{-3}$ (Jenkins \& Tripp, 2001).
The pressure discrepancy between the LB and local warm clouds could
be alleviated somewhat if half of the soft X-ray flux previously
attributed to the LB is actually due to SWCX ions.  Since
the flux is proportional to $n_e^2$, the
thermal pressure is proportional to the square root of the flux.
If the LB flux were
reduced by a factor of 2, as suggested by Cravens (2000) and
Robertson \& Cravens (2003),
the LB pressure would decrease by a
factor of $\sqrt{2}$, but this is not sufficient to explain the
large pressure discrepancy between the LB and 
the local warm cloud pressures.

  \item
On the other hand, 
we do not need to invoke SWCX X-rays in order to lower the LB thermal pressure
which could be smaller than the estimated 15,000~K~cm$^{-3}$
if the gas is far from CIE. For example,
Breitschwerdt (2001) explained the LB's soft X-ray flux as
emission from a recombining plasma whose $P_{th}/k$ was
only $\sim2000$~K cm$^{-3}$.

  \item
Alternatively, a high thermal pressure in the Local Bubble may well
be correct; it may be needed to balance the pressure of
the material above the LB
($P_{th}/k \sim 20,000$~K~cm$^{-3}$, Boulares \& Cox 1990).

\item
Lastly, it is possible that there is not sufficient pressure
support to balance the weight of overlying disk and halo material.
Some observations show material moving downwards at
$\sim 20$~km~s$^{-1}$.

\end{enumerate}

\item
If one concludes that the LB does not exist, 
then one must find some other material
to fill the space within the Local Cavity and provide
pressure to balance the warm clouds and the overlying material.
Suggestions for solving this problem include:
\begin{enumerate}
\item
Diffuse material having a high magnetic field could provide sufficient
magnetic pressure to support the overlaying material.    
Andersson \& Potter (2006) found that the magnetic
field strength in the plane of the sky is 
$8^{+5}_{-3}\ \mu$G,
creating a magnetic pressure of $P_B/k \sim 18,000$~K~cm$^{-3}$ 
on sightlines terminating within about 200~pc of the Sun. 
However, in his conference presentation, Steve Spangler reported 
a LB sight line
(toward pulsar J0437-4715, located 170~pc from the Sun), 
for which the interstellar magnetic field strength is $0.7\ \mu$G.   
The magnetic
pressure along that sight line would be far too small to balance the pressure
of the adjacent gas.

\item
Photoionized gas with a temperature of $T \sim 20,000$~K
could fill the space, but such gas may be disallowed by observations.
\item
We also considered turbulent pressure and ram pressure. 
We abandoned turbulent pressure because it does
not provide enough support (Redfield \& Linsky 2004), but    
ram pressure could provide significant support (de Avillez 2007).
\end{enumerate}
\item
Is the Local Bubble expanding? Probably not.
If anything, the Loop I Bubble is expanding into the Local Bubble.
\end{enumerate}

\subsection{Energy Balance} 

\begin{enumerate}
\item
The hot gas in the LB is cooling, but it's energy loss is not balanced by 
energy gain. 
\item
The embedded warm clouds are also losing energy by radiation, but they
may also gain energy via
photoionization by the extreme ultraviolet and
soft X-ray photons from the LB (see Slavin \& Frisch, 2002). 
\item
We considered the possibility that heat may be conducted into the
clouds from the surrounding hot LB plasma (see Slavin 1989). 
\begin{enumerate}
 \item
However, cloud interfaces may be complicated and the conditions on either
side of the interface are not well known.  
 \item
If the cloud is surrounded by a tangled 
magnetic field, then the magnetic field will quench thermal
conduction.
 \item
Some lines of sight through embedded clouds do not contain
observable numbers of \oxysix\ ions (a tracer of evaporative
interfaces), but this may be result from the suppression of
thermal conduction by magnetic fields
and, therefore, cannot be taken as evidence
against the the existence of hot gas in the Local Bubble.
\end{enumerate}
\end{enumerate}

\section{ Small Scale Structure in the LB } 
\label{smallstructure}

\begin{enumerate}
\item
Warm ($T \sim 8000$~K), parsec-scale clouds exist within the LB (Redfield
\& Linsky 2008). 
\item
One explanation for the presence of such clouds within the hostile 
environment of the Local Bubble is the Cox \& Helenius (2003) model, 
in which the magnetic field drags material from the LB wall into
the interior of the bubble.
\item
In addition to the population of parsec-scale clouds,
there may also be a population of smaller clouds.
Scintillation observed in QSO signals implies that
very small structures exist in very local ISM (Linsky et al. 2008). 
\item
The very small scale clouds (diameter = 800 to 4000~AU) that exist
in the general interstellar medium may also exist within the
Local Bubble (see Snezana Stanimirovic's paper in this volume).
They could be neutral or ionized and may be fragments  
from the LB wall, as in the Cox \& Helenius model or they may
result from collisions of the parsec-scale clouds (Redfield 2007)
\item
However, evaporation should destroy clouds on Myr timescales unless
the clouds are well protected by the magnetic field. 
\end{enumerate}

\section{ Other Important Physical Processes? } 
\label{otherprocesses}

\begin{enumerate}
\item
The standard assumption used when analyzing the X-ray data has been
that the gas is in CIE but this 
assumption should be re-examined. 
\item
Breitschwerdt \& Schmutzler (1994)
and Breitschwerdt (2001) computed
drastically underionized models of the LB which explained the 
observed quantity of soft X-ray photons. 
\item
The Breitschwerdt \& Schmutzler scenario 
is disallowed because the ratios of the observed 
X-rays to \oxysix\ and \carthree\ ions and intensity
are higher than expected from the model
(Welsh et al. 2002; Shelton 2003; Oegerle et al. 2005). 
\item
In order for the Breitschwerdt \& Schmutzler scenario to be valid,
the LB's soft X-ray intensity would need to be reduced by a large factor.
\end{enumerate}

\vspace{0.5cm}
\noindent
{\bf{Acknowledgments}}

I would like to thank the conference conveners 
(Dieter Breitschwerdt, Priscilla Frisch,
Vlad Izmodenov, Jeffrey Linsky, and Eberhard M\"{o}bius)
for organizing an extremely productive and enjoyable 
conference.   
Special thanks go
to Jeffrey Linsky, who ably facilitated the discussions in Sessions 1 
and 3, asked me to write up this session report for the
conference proceedings, and improved the manuscript.
It is a pleasure to credit my fellow members of Sessions 1
and 3 
(Miguel de Avillez, 
Dieter Breitschwerdt, 
Berkhard Fuchs,
Ed Jenkins, 
Dimitra Koutroumpa,
Rosine Lallement, 
Steve Snowden,
Steve Spangler, 
Snezana Stanimirovic, and
Barry Welsh) 
for providing the discussions recorded in this
report.
In the question and answer periods during and after the session
report was orally delivered, the larger group of workshop participants
refined and supplemented some of our thinking.   
Brian Wood and Seth Redfield added ideas that have been incorporated
into this manuscript.
In addition, I would like to acknowledge the generosity of 
the International Space Science Institute (ISSI) and
NASA through LTSA grant NNG04GD78G for funding my trip to the
conference.

\end{document}